\documentclass[aps, prx, groupaddress, reprint, showpacs, twocolumn]{revtex4-1}

\usepackage{amsfonts}
\usepackage{amsmath}
\usepackage{amssymb, bm, mathrsfs}
\usepackage{graphicx, epstopdf, color}
\usepackage{soul}
\usepackage[colorlinks=true,citecolor=blue, urlcolor  = blue,]{hyperref}
\usepackage{sidecap}

\begin{document}
	
	\title{\textcolor{blue}{Publisher link: \href{https://doi.org/10.1103/PhysRevApplied.12.014057}{Phys. Rev. Appl. 12, 014057(2019)}} \\Generalized High-Energy Thermionic Electron Injection at Graphene Interface} 

	\author{Yee Sin Ang}
	\email{yeesin\_ang@sutd.edu.sg}
	\affiliation{Science and Mathematics, Singapore University of Technology and Design, 8 Somapah Road, Singapore 487372}

	\author{Yueyi Chen}
	\affiliation{Science and Mathematics, Singapore University of Technology and Design, 8 Somapah Road, Singapore 487372}
	
	\author{Chuan Tan}
	\affiliation{Science and Mathematics, Singapore University of Technology and Design, 8 Somapah Road, Singapore 487372}

	\author{L. K. Ang}
	\email{ricky\_ang@sutd.edu.sg}
	\affiliation{Science and Mathematics, Singapore University of Technology and Design, 8 Somapah Road, Singapore 487372}

\begin{abstract}

Graphene thermionic electron emission across high-interface-barrier involves energetic electrons residing far away from the Dirac point where the Dirac cone approximation of the band structure breaks down. 
Here we construct a full-band model beyond the simple Dirac cone approximation for the thermionic injection of high-energy electrons in graphene.
We show that the thermionic emission model based on the Dirac cone approximation is valid only in the graphene/semiconductor Schottky interface operating near room temperature, but breaks down in the cases involving high-energy electrons, such as graphene/vacuum interface or heterojunction in the presence of photon absorption, where the full-band model is required to account for the band structure nonlinearity at high electron energy.
We identify a critical barrier height, $\Phi_B^{(\text{c})} \approx 3.5$ eV, beyond which the Dirac cone approximation crosses over from underestimation to overestimation.
In the high-temperature thermionic emission regime at graphene/vacuum interface, the Dirac cone approximation severely overestimates the electrical and heat current densities by more than 50\% compared to the more accurate full-band model. 
The large discrepancies between the two models are demonstrated using a graphene-based thermionic cooler. 
These findings reveal the fallacy of Dirac cone approximation in the thermionic injection of high-energy electrons in graphene. 
The full-band model developed here can be readily generalized to other 2D materials, and shall provide an improved theoretical avenue for the accurate analysis, modeling and design of graphene-based thermionic energy devices.

\end{abstract}

\maketitle	

\section{Introduction}

Recent theoretical and experimental developments have revealed graphene's extraordinary potential in various thermionic-based energy applications \cite{yuan, massicotte}. 
In thermionic emission process, electrons are thermally excited to overcome the interface potential barrier and emitted across the interface. Collection of these emitted electrons at the anode forms an electricity through an external load, thus achieving thermionic-based heat-to-electricity conversion.
Recent experiment has demonstrated graphene monolayer as a highly efficient anode for direct heat-to-electricity energy conversion with high conversion efficiency reaching 9.8\%, which can be further optimized by electrostatic gating and cathode-anode gap reduction \cite{yuan}.
Thermionic emission of electrons over an insulating barrier can also be harnessed to achieve electronic cooling effect \cite{shakouri, mahan, shakouri2, vanshee}. 
The performance is, however, fundamentally limited by thermal backflow directed from hot to cold electrodes. 
Recent advancements of 2D material van der Waals heterostructure \cite{geim, novoselov, liu} (VDWH) offer new opportunities in solid-state thermionic cooler. 
The layered structure of VDWH strongly impedes phonon propagation and effectively diminishes the thermal backflow effect that is detrimental to the efficiency of thermionic cooler \cite{liang, wang, wang2}. Photon-enhanced thermionic injection across graphene/WSe$_2$/graphene VDWH represents another novel route towards the efficient broadband photodetection and harvesting of light energy in a compact solid-state platform \cite{massicotte}.

\begin{figure}[b]
	\includegraphics[scale=0.3885]{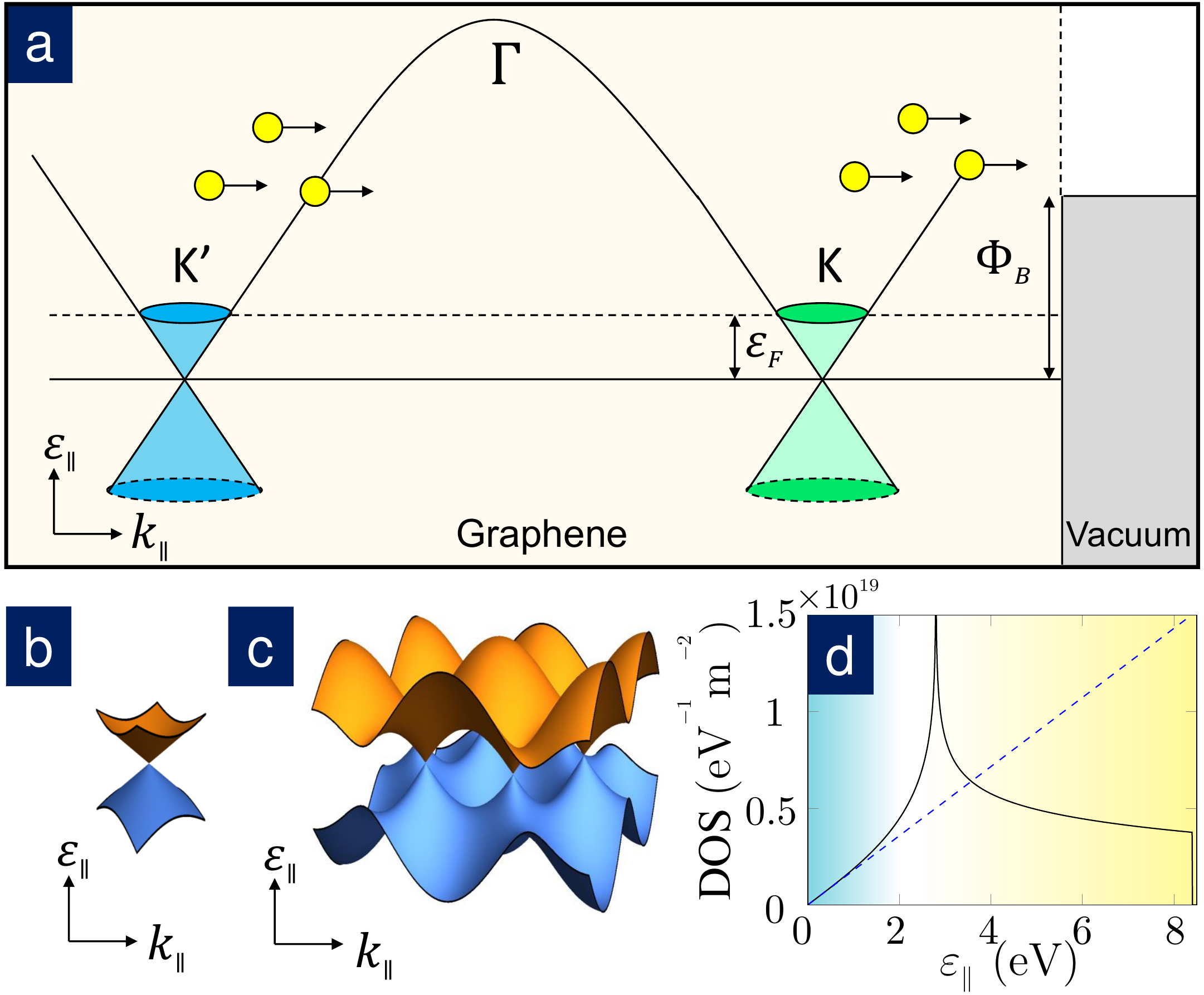}
	\caption{Model of thermionic emission in graphene. (a) Band diagram showing the thermionic electron emission based on the full-band model; (b) Dirac cone at $K$ and $K'$ valley of the first Brillouin zone; (c) full energy band of graphene; and (d) electronic density of states (DOS) showing linear Dirac approximation model (dashed) and full-band model (solid). }
\end{figure}

Previous theoretical models \cite{nieva, misra, misra2, zhang, zhang2, SL, liang2, ang, ang2} describes the out-of-plane thermionic electron emission from the surface of graphene via two key assumptions: 
(i) electron wavevector component lying in the plane of graphene, denoted as $k_\parallel$, is conserved during the out-of-plane emission process \cite{liang2}; and
(ii) electrons undergoing thermionic emission are described by a conic energy band structure, commonly known as the \emph{Dirac cone approximation} \cite{SL, liang2, ang3,  trushin, trushin2}. 
These assumptions break down in the case of thermionic emission of high-energy electrons due to the following reasons. 
Firstly, the conservation of $\varepsilon_\parallel$ is violated due to the presence of electron-electron, electron-impurity, and electron-phonon scattering effect \cite{meshkov, russell, perebeinos, vdovin, liu_acs}. Clear experimental evidence of $k_\parallel$-nonconserving vertical electron transport due to electron-phonon scattering has been observed in graphene heterostructures \cite{vdovin}. 
At typical device operated at room temperature and above, electrons in graphene are expected to undergo strong scatterings with phonons and impurities. 
Thus, at the high-temperature thermionic emission regime, $k_\parallel$-conservation is expected to be strongly violated due to phonon \cite{vdovin} and impurity scattering \cite{liu_acs} effects, which immediately implies the breakdown of assumption (i).
Secondly, as the interface potential exceeds $\Phi_B \approx 1$ eV, the thermionic emission should be dominantly contributed by energetic carriers with $\varepsilon_\parallel > 1$ eV, where the band structure deviates significantly from the simple Dirac cone approximation. 
Barrier-lowering mechanisms, such as surface coating which lowers the vacuum level by about 1 eV, and electrostatic doping via gate voltage which raises the Fermi level by up to 1 eV \cite{voss, yu, yuan2}, can typically reduce $\Phi_B$ by up to $\sim 1$ eV. 
In this case, electrons undergoing thermionic emission are still energetically beyond the Dirac cone regime.
Furthermore, in the presence of photon absorption \cite{massicotte,misra2}, electrons are energetically promoted to high-energy state beyond the Dirac cone regime.
The Dirac cone approximation is thus expected to produce inaccurate results for high-temperature, high-barrier and photon-enhanced thermionic injections in graphene-based heterointerface. For this reason, the assumption (ii) should also be revised to account for the band structure nonlinearity at high energy.

In this paper, we construct an improved graphene thermionic emission model that relaxes the two assumptions discussed above by explicitly including the effects of: (i) $k_\parallel$-nonconservation \cite{ang3}; and (ii) band structure nonlinearity beyond the simple Dirac cone approximation by using the full-band tight-binding energy dispersion \cite{reich}. 
We found that the simple Dirac-based model is only valid for solid-state graphene/semiconductor Schottky contact with low interface-barrier ($\Phi_B < 1$ eV) \cite{tongay} operating near room temperature ($T \approx 300$ K).
At high-barrier ($\Phi_B \gg 1$ eV) and high-temperature ($T >1000$ K) regimes, the Dirac-based model deviates significantly from the full-band model developed here, thus revealing the shortcoming of Dirac cone approximation in the thermionic emission of graphene/vacuum interface.
We identify the existence of a \emph{critical potential barrier height}, $\Phi_B^{(c)} \approx 3.5$ eV, beyond which the Dirac cone approximation crosses over from underestimation to overestimation in comparison with the more accurate full-band model. 
We calculate the efficiency of a graphene-based thermionic cooler device using both Dirac approximation and the improved full-band model, and a large discrepancy between the two models is revealed. These findings prompt an urgent need in replacing the Dirac cone approximation with the full-band model developed here so to achieve a more reliable modeling and understanding of graphene-based high-temperature thermionic energy devices.

\section{Theory}

The thermionic electrical and heat current densities from the surface of a 2D electronic system are [see Fig. 1(a) for the energy band diagram],
\begin{subequations}
\begin{equation}\label{current}
\mathcal{J} = \frac{g_{s,v} e}{(2\pi)^2} \sum_i \frac{v_\perp^{(i)}(k_\perp^{(i)})}{L_\perp} \int \text{d}^2\mathbf{k}_\parallel \mathcal{T}^{(i)}(\mathbf{k}_\parallel, k_\perp) f(\varepsilon_{\mathbf{k}_\parallel}),
\end{equation}
\begin{eqnarray}\label{heat}
\mathcal{Q} &=& \frac{g_{s,v} }{(2\pi)^2} \sum_i \frac{v_\perp^{(i)}(k_\perp^{(i)})}{L_\perp} \nonumber \\
&& \times  \int \text{d}^2\mathbf{k}_\parallel (\varepsilon_\parallel -\varepsilon_F) \mathcal{T}^{(i)}(\mathbf{k}_\parallel, k_\perp) f(\varepsilon_{\mathbf{k}_\parallel}),
\end{eqnarray}
\end{subequations}
where $g_{s,v}=4$ is the spin-valley degeneracy, $L_\perp$ is the 2D material thickness ($L_\perp = 0.335$ nm for graphene \cite{ni}), $\varepsilon_F$ is the Fermi level, $f(\varepsilon_{\mathbf{k}_\parallel})$ is the Fermi-Dirac distribution function, $\mathbf{k}_\parallel = (k_x, k_y)$ is the electron wave vector component lying in the 2D plane, $\perp$ denotes the direction orthogonal to the $x$-$y$ plane of the 2D system, $k_\perp^{(i)}$ is the quantized out-of-plane wave vector component of $i$-th subband, $v_\perp^{(i)}(k_\perp^{(i)}) = \sqrt{2m\varepsilon_\perp^{(i)}}$ is the cross-plane electron group velocity, $m$ is the free electron mass, $\varepsilon_\perp^{(i)}$ is the discrete bound state energy level, and the summation, $\sum_i \cdots$, runs over all of the $i$-th quantized subbands. 
The non-conservation of $\mathbf{k}_\parallel$ during the out-of-plane thermionic emission process leads to the coupling between $\mathbf{k}_\parallel$ and $k_\perp$. Accordingly, the $i$-th subband transmission probability becomes $\mathcal{T}^{(i)}(k_\perp, \mathbf{k}_\parallel)$, i.e. the cross-plane electron tunneling is dependent on both $k_\perp$ and $\mathbf{k}_\parallel$. The $\mathbf{k}_\parallel$-nonconserving model has been extensively studied in previous theoretical works \cite{vanshee, ang3, vashaee2, dwyer, kim} and has been successfully employed in the analysis of thermionic transport experiments in graphene-based devices \cite{massicotte, ma, SL}.

For thermionic emission, the transmission probability can be written as $\mathcal{T}^{(i)}(k_\perp, \mathbf{k}_\parallel) = \lambda \Theta\left( \varepsilon_\parallel + \varepsilon_\perp^{(i)}- \Phi_{B} \right)$, i.e. $\varepsilon^{(i)}_\perp$ and $\varepsilon_\parallel$ are combined to overcome the interface barrier $\Phi_{B}$. Here $\lambda$ is a parameter representing the strength of $\mathbf{k}_\parallel$-non-conserving scattering processes. The term $\Theta(x)$ denotes the Heaviside step-function. 
Equations (\ref{current}) and (\ref{heat}) can be simplified into a single-subband thermionic emission electrical and heat current densities for graphene, respectively, as
\begin{subequations}\label{flux}
\begin{equation}\label{current_sim}
\mathcal{J} = \frac{g_{s,v}ev_\perp}{(2\pi)^2} \int \text{d}^2 \mathbf{k}_\parallel \mathcal{T}(\varepsilon_\parallel) \xi_T(\varepsilon_\parallel),
\end{equation}
\begin{equation}\label{heat_sim}
\mathcal{Q} = \frac{g_{s,v}v_\perp}{(2\pi)^2} \int \text{d}^2 \mathbf{k}_\parallel (\varepsilon_\parallel - \varepsilon_F) \mathcal{T}(\varepsilon_\parallel) \xi_T(\varepsilon_\parallel),
\end{equation}
\end{subequations}
where $\xi_T(x) \equiv \exp\left( -\frac{x - \varepsilon_F}{k_BT} \right)$. In writing $\mathcal{T}(\varepsilon_\parallel) = \lambda \Theta\left(\varepsilon_\parallel(\mathbf{k}_\parallel) - \Phi_{B}\right)$, we have used the fact that the quantized subband energy level, $\varepsilon_\perp^{(1)}$, can be absorbed into $\varepsilon_\parallel$ by setting $\varepsilon_\perp^{(1)}$ as the zero-reference of $\varepsilon_\parallel$ \cite{ang3}, and denoted $v_\perp \equiv v_\perp^{(1)}(k_\perp^{(1)})$. 
The Fermi-Dirac distribution function approaches the semiclassical Maxwell-Boltzmann distribution function, $\xi_T(\varepsilon_\parallel)$, since the emitted electrons are in the non-degenerate regime, i.e. $\varepsilon_\parallel \approx \Phi_{B}$, and $\Phi_{B} \gg \varepsilon_F$ for typical values of $\varepsilon_F < 1$ eV and graphene/vacuum interface barrier of $\Phi_{B} = 4.5$ eV.

The electronic properties of graphene enter Eq. (\ref{current_sim}) via the $\mathbf{k}_\parallel$-integration. 
The $\mathbf{k}_\parallel$-integral is transformed rewritten as $\text{d}^2 \mathbf{k}_\parallel/(2\pi)^2 = D(\varepsilon_\parallel) d\varepsilon_\parallel$, where $D(\varepsilon_\parallel)$ is the electronic density of states (DOS). In general, the $\mathbf{k}_\parallel$-nonconserving thermionic emission model in Eq. (\ref{flux}) can be solved for any 2D materials using the appropriate DOS. Here we shall focus on solving Eq. (\ref{flux}) for graphene. 
The full-band tight-binding model of graphene yields,  
\begin{equation}\label{full_DOS}
D_{\text{FB}}(\varepsilon_\parallel) = 
\begin{cases}
\frac{D_0}{\sqrt{F\left(\varepsilon_\parallel/t'\right)}} \mathbb{K} \left( \frac{4\varepsilon_\parallel/t'}{F\left(\varepsilon_\parallel/t'\right)} \right) & 0 <\varepsilon_\parallel < t' \\
\frac{D_0}{\sqrt{4\varepsilon_\parallel/t'}} \mathbb{K} \left( \frac{F\left(\varepsilon_\parallel/t'\right)}{ 4\varepsilon_\parallel/t' } \right) & t' <\varepsilon_\parallel < 3t' 
\end{cases},
\end{equation}
where $D_0 \equiv \frac{1}{A_c} \frac{g_{s,v}}{\pi^2} \frac{\varepsilon_\parallel}{t'^2}$, $t' = 2.8$ eV, $a = 0.142 $ nm, $A_c = 3a^2\sqrt{3}/2$, $F(x) \equiv (1+x)^2 - (x^2 - 1)^2/4$, and $\mathbb{K}(m) \equiv \int_0^1 dx \left[ (1-x^2)(1-mx^2) \right]^{-1/2}$ is the complete Elliptic integral of the first kind \cite{abramowitz}.

The conduction and valence band touches at the $K$ and $K'$ points in the first Brillouin zone, commonly known as the \emph{Dirac cone} [Figs. 1(b) and (c)]. 
At the vicinity of Dirac cone, $\varepsilon_\parallel$ can be expanded up to the first order in $|\mathbf{k}_\parallel|$ to yields a pseudo-relativistic relation, $\varepsilon_\parallel = \hbar v_F |\mathbf{k}_\parallel|$, where $v_F = 10^6$ m/s.
The corresponding DOS is
\begin{equation}\label{DOS_Dirac}
D_{\text{D}}(\varepsilon_\parallel) = \frac{g_{s,v} \varepsilon_\parallel}{2\pi\hbar^2 v_F^2},
\end{equation}
which exhibits a monotonous linear relation with $\varepsilon_\parallel$. 
This linear energy dispersion and the corresponding density of states have led to many unusual physical phenomena in graphene, such as Klein tunnelling effect \cite{klein}, room-temperature quantum Hall effect \cite{hall}, exceptionally large electron mobility {mobility}, gate-tunable optical and plasmonic responses \cite{plasmonics}, strong optical nonlinearity \cite{NOR} and the emergence of new electromagnetic modes \cite{em}. 
It should be noted that the $D_{\text{D}}(\varepsilon_\parallel)$ is in good agreement with $D_{\text{FB}}(\varepsilon_\parallel)$ only for $\varepsilon_\parallel < 1$ eV. 
For $\varepsilon_\parallel > 1$ eV, the ever-increasing $D_{\text{D}}(\varepsilon_\parallel)$ severely overestimates the actual DOS calculated from the full-band model [see Fig. 1(d)]. 
The question of how the high-energy discrepancy between $D_{\text{D}}(\varepsilon_\parallel)$ and $D_{\text{FB}}(\varepsilon_\parallel)$ affect the thermionic emission of energetic electrons with $\varepsilon_\parallel > 1$ eV remains unanswered thus far.  
As demonstrated below, we found that the band nonlinearity effect at high electron energy directly leads to the large discrepancy between the Dirac cone approximation and the full-band thermionic emission model of energetic electrons with $\varepsilon_\parallel > 1$ eV.

By combining Eqs. (\ref{flux}) and (\ref{DOS_Dirac}), the electrical and heat current under Dirac cone approximation can be analytically solved as,
\begin{subequations}\label{dirac}
	\begin{eqnarray}\label{Jd}
	\mathcal{J}_{ \text{D}} &=&  \frac{ \lambda v_\perp}{L_\perp}  \frac{g_{s,v}e\left(k_BT\right)^2}{2\pi\hbar^2v_F^2}  \left( 1 + \frac{\Phi_{B}}{k_BT} \right) \xi_T(\Phi_B),
	\end{eqnarray}
	\begin{eqnarray}
	\mathcal{Q}_{\text{D}} &=& \frac{ \lambda v_\perp }{L_\perp} \frac{g_{s,v} (k_BT)^3}{2\pi\hbar^2 v_F^2} \Lambda
	\xi_T(\Phi_B),
	\end{eqnarray}
\end{subequations}
where $\Lambda \equiv \left( \Phi_{B}/k_BT\right)^2 + \left(2 - \varepsilon_F / k_BT \right) \left(1 + \Phi_{B} / k_BT\right)$.
On the other hand, using the DOS in Eq. (\ref{full_DOS}), the full-band equivalence of Eq. (\ref{dirac}) is
	\begin{subequations}\label{J}
			\begin{equation}\label{phi_small}
			\mathcal{J}_{ \text{FB}}= \lambda \frac{v_\perp}{L_\perp} \frac{g_{s,v}e}{\pi^2 t'^2} \mathcal{I}_0(\Phi_B),
			\end{equation}
			\begin{equation}\label{phi_big}
				\mathcal{Q}_{ \text{FB}}= \lambda \frac{v_\perp}{L_\perp} \frac{g_{s,v}}{\pi^2 t'^2} \mathcal{I}_1(\Phi_B),
			\end{equation}
	\end{subequations}
where $\mathcal{I}_\mu(\Phi_B)$ can be numerically solved from
\begin{widetext}
	\begin{equation}\label{integral}
	\mathcal{I}_\mu(\Phi_B) \equiv \Theta\left( t'-\Phi_B \right)\int_{\Phi_B}^{t'}  \frac{ \varepsilon_\parallel (\varepsilon_\parallel - \varepsilon_F)^\mu d\varepsilon_\parallel }{\sqrt{F\left(\varepsilon_\parallel/t'\right)}} \mathbb{K} \left( \frac{4\varepsilon_\parallel/t'}{  F\left(\varepsilon_\parallel/t'\right)} \right) \xi_T(\varepsilon_\parallel) + \int_{t}^{3t'}  \frac{ \varepsilon_\parallel (\varepsilon_\parallel - \varepsilon_F)^\mu d\varepsilon_\parallel }{\sqrt{4\varepsilon_\parallel/t'}} \mathbb{K} \left( \frac{F\left(\varepsilon_\parallel/t'\right)}{ 4\varepsilon_\parallel/t' } \right)\xi_T(\varepsilon_\parallel) .
	\end{equation}
\end{widetext}
The second term of Eq. (\ref{integral}) is set to $3t'$ as $D_{\text{Gr}}(\varepsilon_\parallel)$ reaches the maximum of the conduction band at $\varepsilon_\parallel \leq3t'$. 
It should be noted that $\mathcal{J}_{\text{D}}$ in Eq. (\ref{Jd}) has been rigorously studied in graphene-based Schottky contacts and a good agreement between theory and experimental data is demonstrated \cite{ma,SL}. 
Such good agreement immediately suggests the need to extend Eq. (\ref{dirac}) via the full-band model in Eq. (\ref{full_DOS}), so to obtain a \emph{generalized} theoretical framework that encompasses both the low-energy thermionic emission in graphene-based Schottky contact and the high-energy counterpart in graphene/vacuum interface.

\section{Results and Discussions}

\begin{figure*}
	\includegraphics[scale=0.85]{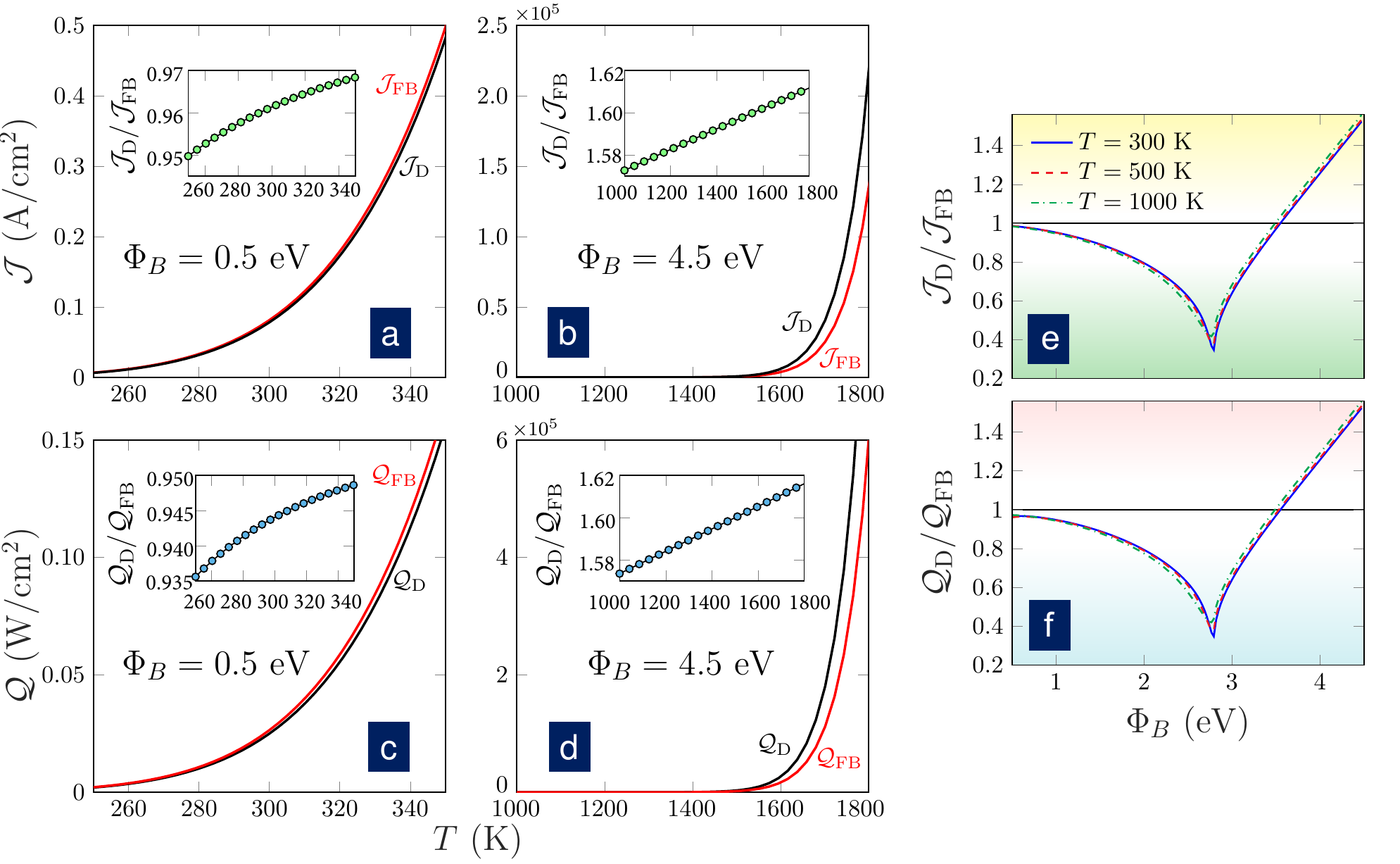}
	\caption{Comparison of Dirac approximation and full-band thermionic models. (a) and (b) shows the electrical and heat current densities at $\Phi_B = 0.5$ eV, respectively. (c) and (d) same as (a) and (b), with $\Phi_B = 4.5$ eV. Insets show the ratios of $\mathcal{J}_{\text{D}}/\mathcal{J}_{\text{FD}}$ and $\mathcal{Q}_{\text{D}}/\mathcal{Q}_{\text{FD}}$. The $\Phi_B$-dependence of (e) $\mathcal{J}_{\text{D}}/\mathcal{J}_{\text{FD}}$; and (f) $\mathcal{Q}_{\text{D}}/\mathcal{Q}_{\text{FD}}$ at different temperatures, $T = 300, 800, 1000$ K. }
\end{figure*}

In Fig. 2, the analytical results of the Dirac cone approximation, $\mathcal{J}_{\text{D}}$ and $\mathcal{Q}_{\text{D}}$, and the numerical results of the full-band model, $\mathcal{J}_{\text{FB}}$ and $\mathcal{Q}_{\text{FB}}$, are plotted for two values of $\Phi_{B}$ which are typical for graphene-based Schottky contact \cite{SL, tongay} and for graphene/vacuum thermionic emitter \cite{liang2}: (i) the low-barrier graphene/semiconductor Schottky diode regime ($\Phi_{B} = 0.5$ eV) around room temperature; and (ii) the high-barrier regime ($\Phi_{B} = 4.5$ eV) operating at $T>1000$ K. 
We have used $v_\perp = 3.7 \times 10^6$ m/s and $L_\perp = 0.335$ nm for graphene \cite{ang3}, and scattering strength \cite{russell} of $\lambda = 10^{-4}$.
In the low-barrier regime, both Dirac and full-band models produce nearly identical electrical [Fig. 2(a)] and heat [Fig. 2(c)] current densities in the typical room temperature operating regime for a Schottky diode. The Dirac model slightly underestimates the electrical and heat current densities by approximately 5\% [see insets of Figs. 2(a) and (c)]. 
Conversely, in the high-barrier graphene/vacuum regime, the Dirac model severely overestimates the electrical and heat current densities by $\sim 60\%$ in the high-temperature range from 1000 K to 1800 K [see Figs. 2(b) and (d)]. 
This rather sizable discrepancy immediately reveals the incompatibility of Dirac cone approximation in the thermionic emission of high-energy electrons occurring at graphene/vacuum interface \cite{liang, wang, wang2} or in the presence of photon absorption \cite{massicotte,ma}.
This fallacy of Dirac cone approximation arises from the fact that graphene band structure becomes highly nonlinear at high electron energy $\varepsilon_\parallel>1$ eV at which the Dirac cone approximation fails to capture this band nonlinearity.  
We thus arrive at the following key finding: the simple analytical Dirac model in Eq. (\ref{dirac}) is only well-suited for the modeling of thermionic emission mediated by electron with energy $< 1$ eV, such as graphene-based Schottky diode \cite{tongay,SL}, while the full-band model in Eq. (\ref{J}) must be used for the thermionic emission of energetic electrons ($> 1$ eV) in high-barrier graphene interface and photon-enhanced thermionic devices so to produce a more accurate modeling results. 

In Figs. 2(e) and (f), we further investigate the difference between the Dirac cone approximation and the full-band model in the intermediate regime between $\Phi_B = 0.5$ eV and $\Phi_B = 4.5$ eV. 
In general, the electrical and heat current density ratios, i.e. $\mathcal{J}_{\text{D}}/\mathcal{J}_{\text{FB}}$ [Fig. 2(e)] and $\mathcal{Q}_{\text{D}}/\mathcal{Q}_{\text{FB}}$ [Fig. 2(f)], are weakly dependent on temperature, and both ratios are approximately equal, i.e. $\mathcal{J}_{\text{D}}/\mathcal{J}_{\text{FB}} \approx \mathcal{Q}_{\text{D}}/\mathcal{Q}_{\text{FB}}$. 
For $\Phi_B$ lying approximately between $0.5$ eV and $3.5$ eV, we found that $\mathcal{J}_{\text{D}}/\mathcal{J}_{\text{FB}}\approx\mathcal{Q}_{\text{D}}/\mathcal{Q}_{\text{FB}} < 1$, which signifies the underestimation of the thermionic emission current densities by the Dirac cone approximation in comparison with the full-band model. 
Such underestimation peaks at $\Phi_B \approx 2.7$ eV with $\mathcal{J}_{\text{D}}/\mathcal{J}_{\text{FB}} \approx \mathcal{Q}_{\text{D}}/\mathcal{Q}_{\text{FB}} \approx 0.35$. 
Interestingly, there exists a critical barrier height, $\Phi_B^{(\text{c})} \approx 3.5$ eV, beyond which the current density ratios switch from $\mathcal{J}_{\text{D}}/\mathcal{J}_{\text{FB}} \approx \mathcal{Q}_{\text{D}}/\mathcal{Q}_{\text{FB}} < 1$ to $\mathcal{J}_{\text{D}}/\mathcal{J}_{\text{FB}} \approx \mathcal{Q}_{\text{D}}/\mathcal{Q}_{\text{FB}} > 1$, thus signifying the transition from underestimation to overestimation of the thermionic emission due to the Dirac cone approximation.
This critical $\Phi_B^{(\text{c})}$ corresponds to an \emph{accidental DOS averaging effect} at which the underestimation of the electron population available for thermionic emission due to the Dirac cone approximation at energy slightly above $\Phi_B$ is \emph{exactly compensated} by the overestimation of that at higher energy. 
Beyond this critical barrier height, the overestimation of the electron population available for thermionic emission due to Dirac cone approximation becomes increasingly severe, which directly leads to the monotonously increasing trend in both $\mathcal{J}_{\text{D}}/\mathcal{J}_{\text{FB}}$ and $\mathcal{Q}_{\text{D}}/\mathcal{Q}_{\text{FB}}$ as $\Phi_B > \Phi_B^{(\text{c})}$.

We now investigate the current-temperature scaling of the full-band thermionic emission model. For 2D materials, the current-temperature scaling of thermionic emission in the out-of-plane direction follows the universal scaling law, $\ln{\left(\mathcal{J}/T\right)} \propto -1/T$ \cite{ang3}, rather than the classic Richardson-Dushman scaling law of $\ln{\left(\mathcal{J}/T^2\right)} \propto -1/T$ for 3D materials \cite{p_zhang, shinozaki}. In Fig. 4, the numerical value of $\ln{\left(\mathcal{J}_{\text{FD}}/T\right)}$ is plotted against $1/T$. The numerical results fit excellently into a straight line for both low-barrier [Fig. 3(a)] and high-temperature [Fig. 3(b)] regimes, thus confirming the expected universal scaling behavior in the full-band model of graphene. 

\begin{figure}[t]
	\includegraphics[scale=0.85]{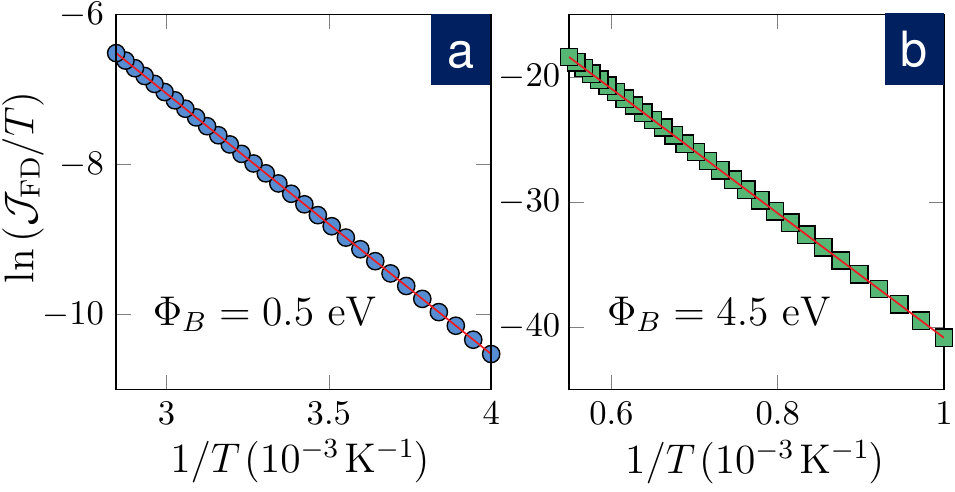}
	\caption{Modified Arrhenius plot of $\ln\left( \mathcal{J}_{\text{FB}}/T \right) \propto -1/T$ for (a) $\Phi_B = 0.5$ eV; and (b) $\Phi_B = 4.5$ eV.}
\end{figure}

To further investigate the impact of the full-band model on the modeling of graphene-based thermionic energy device, we calculate the thermionic cooling efficiency of a graphene thermionic cooler (see inset of Fig. 4(c) for a schematic drawing of the energy band diagram). 
Here, the thermionic cooler is composed of two graphene electrodes in parallel-plate configurations where a bias voltage, $V$, is used to modulate the net emitted electrical and heat current to achieve cooling \cite{liang2}. 
The hot (cold) graphene electrode temperature is denoted as $T_H$ ($T_C$).
In Figs. 4(a) and (b), we see that the Dirac model overestimates both electrical and heat current densities immediately after the onset of cooling effect where $\Delta \mathcal{Q} >0$. 
The coefficient of performance (COP) of the thermionic cooler is calculated as $\eta(V, T_c, T_h) = \Delta Q/V\Delta J$, $\Delta Q = Q(T_c) - Q(T_H)$ and $\Delta \mathcal{J} \equiv \mathcal{J}(T_c) - \mathcal{J}(T_H)$ are the net heat and electrical current densities, respectively. 
In Fig. 4(c), the COP, normalized by Carnot efficiency $\eta_c\equiv T_C/(T_H - T_C)$, is plotted as a function of $V$ with $T_c = 1400$ K, $T_H = 1600$ K and $\varepsilon_F = 0.2$ eV, which reveals an especially large discrepancy between the two models near the onset of cooling at $V \approx 0.65$ V. 
Around the maximal efficiency point $\eta_{\text{max}}$, the full-band model yields $\eta_{\text{max}}$ = 0.75 at 0.7 V, compared to $\eta_{\text{max}}$ = 0.7 at 0.8 V as predicted by the less accurate Dirac model.
These discrepancies can have a significant impact on the practical design of graphene-based thermionic cooler as it can affect multiple values, such as net transported heat current density and the optimal bias voltage, that are crucially important for the optimization of device figures of merit. 
Finally, we remark that the simplistic graphene thermionic cooler model reported in Fig. 4 is aimed to illustrate the discrepancy between the Dirac cone approximation and the full-band model. Realistic modeling of graphene-based thermionic energy device should include important effects, such as image potential lowering \cite{liang2}, blackbody radiation \cite{zhang}, space charge \cite{ang4, p_zhang, shinozaki, ang_QSCL}, electric-field-induced Fermi level shifting \cite{meric}, secondary electron emission \cite{ueda, ueda2}, and carrier scattering effects \cite{meshkov, russell, vdovin, liu_acs}. 
Such detailed modeling is beyond the scope of this work and shall form the subjects of future works. Importantly, the generalized 2D thermionic emission model of graphene developed here shall provide a theoretical foundation that may be directly useful for both the theoretical and experimental studies of the above-mentioned effects.

\begin{figure}[t]
	\includegraphics[scale=1.05]{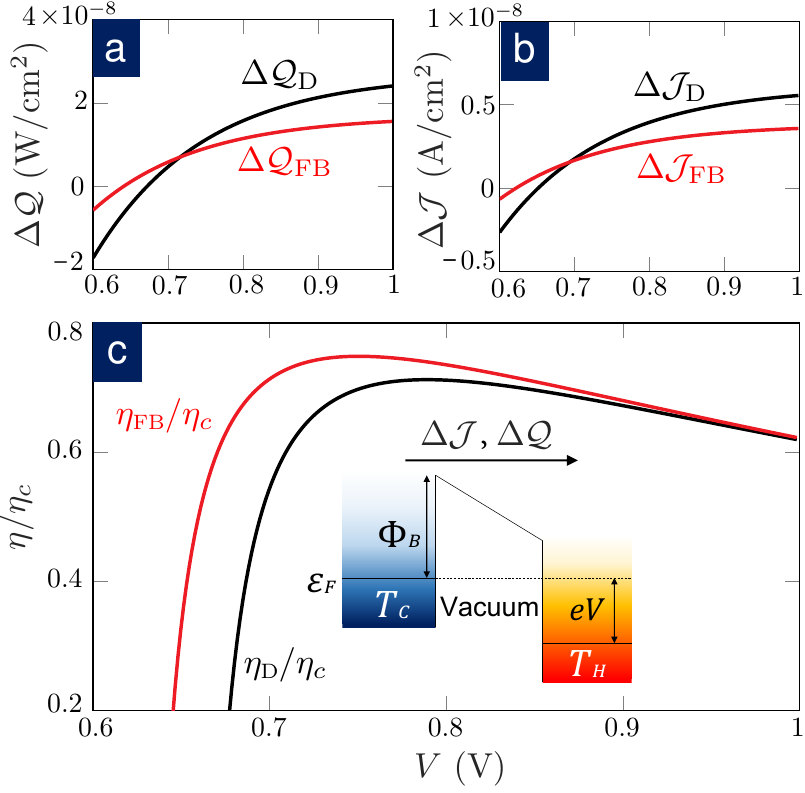}
	\caption{Net (a) heat and (b) electrical current densities, and (c) efficiency of thermionic cooler calculated with $T_H = 1600$ K, $T_C = 1400$ K, and $\varepsilon_F = 0.2$ eV. Inset shows the band diagram of graphene thermionic cooler, composed of cold ($T_C$) and hot ($T_H$) graphene electrodes and biased by, $V$.}
\end{figure}

Finally, for completeness, we compare the model developed above with the widely-used classic Richardson-Dushman (RD) model \cite{RD}, i.e. $\mathcal{J}_{\text{RD}} = \mathcal{A}_{\text{RD}} T^2 \exp\left( - \Phi_B / k_BT \right)$ where $\mathcal{A}_{\text{RD}} \approx 120$ Acm$^{-2}$K$^{-2}$ is the RD constant. 
Due to the strong dominance of the exponential term at the typical operating regime of $\Phi_B \gg k_BT$, the experimental data of graphene thermionic emission could still be fitted using the classic RD model \cite{zhu, starodub}, especially through the RD scaling law  $\ln\left( \mathcal{J}_{\text{RD}} / T^2 \right) \propto -1/T$, without yielding significant errors in the extraction of $\Phi_B$ \cite{ang}. 
However, the \emph{magnitude} of current density can deviate by several orders of magnitude when an inappropriate model is used \cite{ang}. 
Previous experiments have demonstrated that the extracted pre-exponential term can differ by nearly two orders of magnitude between the classic RD model and the 2D graphene thermionic emission model with Dirac cone approximation \cite{SL,liang3}. 
Such large deviation can severely impact applications of the model in cases where the magnitude of the emission current density is important. 
We quantitatively compare the classic RD model and the generalized full-band model developed in this work by defining the ratio, $\mathcal{J}_{\text{RD}} / \mathcal{J}_{\text{FB}}$. For the two cases studied above, i.e. $\Phi_B = 0.5$ eV at $T = 300$ K and $\Phi_B = 4.5$ eV at $T = 1200$ K for high-temperature graphene/vacuum field emitter, we obtain $\mathcal{J}_{\text{RD}} / \mathcal{J}_{\text{FB}} \approx 1.28\times 10^3$ and $\mathcal{J}_{\text{RD}} / \mathcal{J}_{\text{FB}} \approx 11.6$, respectively. 
This exceedingly large ratios of $\mathcal{J}_{\text{RD}} / \mathcal{J}_{\text{FB}} \gg 1$ suggests that the RD model can severely overestimate the thermionic emission current densities in graphene. 
Thus, the generalized model proposed here, which rigorously captures the detailed high-energy features of the band structure and the two-dimensionality of graphene, shall be more advantageous than both the classic RD model and the simplified model based on Dirac cone approximation, especially for purposes where the magnitude of thermionic emission current density is important, such as the modelling, computational design and parametric optimization of graphene-based thermionic devices \cite{liang, misra,misra2,zhang,zhang2,yang_z}. 

\section{Conclusion}

In conclusion, we have revealed the fallacy of Dirac cone approximation in the modeling of high-barrier high-temperature thermionic emission in graphene. While the classic Richardson-Dushman and the Dirac approximation models \cite{SL,liang2, ang3} remain usable for the simple analysis of experimental data \cite{zhu,starodub,liang3}, the full-band model developed here should be used in the case of high-energy thermionic electron emission in graphene. The proposed full-band model is especially critical for the computational design and modeling of graphene-based thermionic energy devices where the magnitude of the emission current densities is required to be determined accurately. As the 2D thermionic emission formalism developed above can be readily generalized to other 2D materials, our findings shall provide an important theoretical foundation for the understanding of thermionic emission physics in 2D materials.

This work is supported by A*STAR AME IRG (A1783c0011) and AFOSR AOARD (FA2386-17-1-4020). Y. C. is supported by SUTD Undergraduate Research Opportunity Program (UROP).

\end{document}